\documentclass[twocolumn,preprintnumbers,superscriptaddress,nofootinbib,aps,prd,floatfix]{revtex4}

\usepackage{enumerate}
\usepackage{amsmath,amssymb}
\usepackage{graphicx}
\usepackage{bbm,slashed}
\usepackage{xspace,slashed}
\usepackage{hyperref}
\usepackage{tablefootnote}
\hypersetup{colorlinks=true, citecolor=blue, urlcolor=blue, linkcolor=blue}
\usepackage[normalem]{ulem}

\newcommand{\mg}{\textsc{MadGraph5\_}aMC@NLO}
\newcommand{\pythia}{\textsc{Pythia8}}

\begin{document}

\providecommand{\abs}[1]{\lvert#1\rvert}

\newcommand{\Znunujets}{(Z\to{\nu\bar{\nu}})+\text{jets}}
\newcommand{\Welnujets}{(W\to{\ell\nu})+\text{jets}}
\newcommand{\Znunujet}{(Z\to{\nu\bar{\nu}})+\text{jet}}
\newcommand{\Welnujet}{(W\to{\ell\nu})+\text{jet}}

\title{Machine-enhanced CP-asymmetries in the electroweak sector}
\begin{abstract}
The violation of charge conjugation (C) and parity (P) symmetries are a requirement for the observed dominance of matter over antimatter in the Universe. As an established effect of beyond the Standard Model physics, this could point towards additional CP violation in the Higgs-gauge sector. The phenomenological footprint of the associated anomalous couplings can be small, and designing measurement strategies with the highest sensitivity is therefore of the utmost importance in order to maximise the discovery potential of the Large Hadron Collider (LHC). There are, however, very few  measurements of CP-sensitive observables in processes that probe the weak-boson self-interactions. In this article, we study the sensitivity to new sources of CP violation for a range of experimentally-accessible electroweak processes, including $W\gamma$ production, $WW$ production via photon fusion, electroweak $Zjj$ production, electroweak $ZZjj$ production, and electroweak $W^\pm W^\pm jj$ production.  We study simple angular observables as well CP-sensitive observables constructed using the outputs of machine-learning (ML) algorithms. We find that the ML-constructed CP-sensitive observables improve the sensitivity to CP-violating effects by up to a factor of five, depending on the process. We also find that inclusive $W\gamma$ and electroweak $Zjj$ production have the potential to set the best possible constraints on certain CP-odd operators in the Higgs-gauge sector of dimension-six effective field theories.
\end{abstract}

\author{Noah Clarke Hall} 
\affiliation{Department of Physics \& Astronomy, University of Manchester, Manchester M13 9PL, UK\\[0.1cm]}
\author{Isaac Criddle} 
\affiliation{Department of Physics \& Astronomy, University of Manchester, Manchester M13 9PL, UK\\[0.1cm]}
\author{Archie Crossland}
\affiliation{Department of Physics \& Astronomy, University of Manchester, Manchester M13 9PL, UK\\[0.1cm]}
\author{Christoph Englert} \email{christoph.englert@glasgow.ac.uk}
\affiliation{School of Physics \& Astronomy, University of Glasgow, Glasgow G12 8QQ, UK\\[0.1cm]}
\author{Patrick Forbes} 
\affiliation{Department of Physics \& Astronomy, University of Manchester, Manchester M13 9PL, UK\\[0.1cm]}
\author{Robert Hankache} 
\affiliation{Department of Physics \& Astronomy, University of Manchester, Manchester M13 9PL, UK\\[0.1cm]}
\author{Andrew D. Pilkington}\email{andrew.pilkington@manchester.ac.uk} 
\affiliation{Department of Physics \& Astronomy, University of Manchester, Manchester M13 9PL, UK\\[0.1cm]}



\maketitle

\section{Introduction}
\label{sec:intro}
Ten years after the Higgs boson discovery at \hbox{CERN~\cite{ATLAS:2012yve,CMS:2012qbp}}, the search for a more complete picture of particle physics than the Standard Model (SM) continues. With mounting pressure on traditionally-motivated new-physics scenarios, measurements and searches increasingly aim to produce model-independent constraints, chiefly through the application of effective field theory methods~\cite{Weinberg:1978kz}. Such approaches imply a plethora of ad hoc anomalous interactions between SM particles and suggest a wide range of possible physics analyses at the Large Hadron Collider (LHC), even when a weak doublet character of the Higgs field is assumed~\cite{Grzadkowski:2010es,Baak:2014ora}. In contrast, a range of astrophysical and cosmological facts have been established that cannot be explained by the SM alone. These highlight particular subsectors of the EFT deformations as particularly motivated BSM candidates for independent investigations. For instance, there is insufficient CP violation in the SM to explain the observed dominance of matter over anti-matter in the Universe, which could indicate the need for additional CP-violating gauge-Higgs interactions. 

The Standard Model Effective Field Theory (SMEFT) extends the SM Lagrangian with CP-odd dimension-six operators, i.e.
\begin{equation}
\label{eq:efflag}
    {\cal{L}}={\cal{L}}_{\text{SM}} + \sum_{i} {c_i\over \Lambda^2} \widetilde{{\cal{O}}}_i\,.
\end{equation}
The Wilson coefficients, $c_i/\Lambda^2$, specify the strength of the anomalous interactions that are induced, and  $\Lambda$ is considered as the cut-off scale for the effective theory. Of particular interest are operators that affect the interactions between the Higgs boson and the electroweak bosons, or the self-interactions of the weak bosons, namely
\begin{equation}
\label{eq:opscp}
\begin{split}
\widetilde{\cal{O}}_{\widetilde{W}}&=\varepsilon_{ijk} \widetilde{W}^{i}_{\mu\nu} W^{j\,\nu\rho} W^{k\,\mu}_{\rho} \,,\\
    \widetilde{\cal{O}}_{\Phi\widetilde{B}}&=\Phi^\dagger\Phi B^{\mu\nu}\widetilde{B}_{\mu\nu}\,,\\
   \widetilde{\cal{O}}_{\Phi \widetilde{W}}&=\Phi^\dagger \Phi  W^{i\,\mu\nu}\widetilde{W}^{i}_{\mu\nu}\,,\\
   \widetilde{\cal{O}}_{\Phi\widetilde{W}B}&=\Phi^\dagger \sigma^i  \widetilde{W}^{i\,\mu\nu}B_{\mu\nu}\,.
\end{split}
\end{equation}
$W^i_{\mu\nu}$ and $B_{\mu\nu}$ are the $SU(2)_L \times U(1)_Y$ field strengths, the tilded quantities refer to the dual field strengths, $\sigma^i$ are the Pauli matrices, and $\Phi$ denotes the Higgs doublet field. These interactions not only fully parameterise additional sources of weak CP-violating gauge-Higgs interactions, but they also form a closed set under the dimension-six renormalisation group flow~\cite{Alonso:2013hga,Grojean:2013kd,Englert:2014cva} so that their constraints form a theoretically consistent subset of CP violation for matching calculations (e.g. these interactions fully parameterise a broad class of lepton extensions of the SM~\cite{Bakshi:2021ofj,Naskar:2022rpg}, for a more general discussion see also~\cite{Degrande:2021zpv}).

The beyond-the-SM matrix-element for a given process following Eq.~\eqref{eq:efflag} is given by
\begin{equation}
\label{eq:diffxsec}
|{\cal{M}}_{\text{BSM}}|^2 = |{\cal{M}}_{\text{SM}}|^2 + 2\hbox{Re}\{{\cal{M}}_{\text{SM}}{\cal{M}}_{\text{d6}}^\ast \}\,+ |{\cal{M}}_{\text{d6}}|^{2}\,.
\end{equation}
The dimension-six amplitude, ${\cal{M}}_{\text{d6}}$, arises from the interactions induced by the operators of Eq.~\eqref{eq:opscp}, whereas the SM amplitude,  ${\cal{M}}_{\text{SM}}$, arises from ${\cal{L}}_{\text{SM}}$. The interference between the SM amplitude and the dimension-six amplitude induces asymmetries in appropriately-constructed (C)P-odd observables. 

In the Higgs sector, searches for new sources of CP-violation have received considerable attention by the LHC experiments in recent years, with particular emphasis on constructing CP-sensitive observables~\cite{ATLAS:2018hxb,ATLAS:2020wny,ATLAS:2021pkb,CMS:2021sdq,ATLAS:2016ifi,CMS:2019jdw,ATLAS:2020evk,CMS:2021nnc}.
Efforts to maximise the sensitivity are necessary and underway, and observables constructed using matrix-element information are part of the existing analysis strategy~\cite{ATLAS:2016ifi,CMS:2019jdw,ATLAS:2020evk,CMS:2021nnc}. Furthermore, machine-learning (ML) approaches have been proposed as an additional or alternative way to improve sensitivity to the asymmetries caused by new sources of CP violation~\cite{Brehmer:2017lrt,Gritsan:2020pib,Bortolato:2020zcg,Barman:2021yfh,Davis:2021tiv,Bhardwaj:2021ujv}. Specifically, these ML approaches can be used to design highly-sensitive physics analyses when the CP-violating effects are difficult to observe. The machine-learning methods are also relatively easy to implement when compared to matrix-element-based techniques.

In comparison to the Higgs sector, surprisingly little attention has been paid to constructing CP-sensitive observables in processes sensitive to weak-boson self-interactions. This is despite the fact that the CP-odd operators in Eq.~\eqref{eq:opscp} induce anomalous weak-boson self-interactions and can be probed using measurements of diboson production and weak-boson fusion/scattering. The ATLAS measurement of a CP-sensitive observable in electroweak $Zjj$ production~\cite{ATLAS:2020nzk} provided world-leading linearised constraints on $c_{\widetilde{W}}/\Lambda^2$ and limits on $c_{\Phi\widetilde{W}B}/\Lambda^2$ that are competitive with those obtained from the Higgs sector. These constraints were derived using a simple angular observable, the rapidity-ordered azimuthal angle between the two jets~(originally proposed in Ref.~\cite{Plehn:2001nj}), without attempting to use the more sophisticated techniques based on matrix-element information or machine-learning algorithms. Similarly, it was recently proposed that measurements of CP-sensitive observables in $W\gamma$ production would provide even better sensitivity to the $c_{\Phi\widetilde{W}B}/\Lambda^2$ operator than was achieved in the electroweak $Zjj$ analysis~\cite{DasBakshi:2020ejz,Biekotter:2021int}. Again, that projection was based on simple angular observables and did not utilise matrix-element information or machine-learning algorithms. 

In this paper, we use machine-learning algorithms to construct CP-sensitive observables for five processes that are sensitive to weak-boson self-interactions: electroweak $Zjj$ production, inclusive $W\gamma$ production, electroweak $W^\pm W^\pm jj$ production, electroweak $ZZjj$ production, and photon-induced $WW$ production ($\gamma\gamma\rightarrow WW$). We have two goals in this study. The primary goal is to motivate new analyses at the LHC experiments in order to address the paucity of measurements of CP-sensitive observables in the electroweak sector. A secondary goal is to test the applicability of the neural-net based method developed in Ref.~\cite{Bhardwaj:2021ujv} when constructing CP-sensitive observables in a wider range of processes. 

The layout of the paper is as follows. In Sec.~\ref{sec:mc}, we give a brief 
introduction of our simulation framework, where we also provide an overview 
of the fiducial search regions used for this study.
After a discussion 
of CP-sensitive observables in Sec.~\ref{sec:cpvar}, we outline the analysis 
selection and limit setting procedure in Sec.~\ref{sec:limit}. 
Section~\ref{sec:results} is devoted to the discussion of our results; we 
summarise and conclude in Sec.~\ref{sec:conc}.

\section{Simulation framework}
\label{sec:mc}
Throughout, we use \mg{}~\cite{Alwall:2014hca} which interfaced with \pythia{} \cite{Sjostrand:2007gs,Sjostrand:2014zea} to simulate events at leading-order precision in QCD. We employ the {\sc{SmEftSim}}~\cite{Brivio:2017btx,Brivio:2020onw} implementation to model the effective interactions of Eq.~\eqref{eq:opscp} via the {\sc{Ufo}}~\cite{Degrande:2011ua} interface. We limit our analysis to interference effects $\sim \hbox{Re}\{{\cal{M}}_{\text{SM}}{\cal{M}}_{\text{d6}}^\ast \}$ of Eq.~\eqref{eq:diffxsec} and the simulated events are produced at $c/\Lambda^2 = 1/\text{TeV}^2$.
``Squared'' CP-even dimension-6 effects $\sim |{\cal{M}}_{\text{d6}}|^{2}$ will not affect the discrimination that we study below and would only change the normalisation of the cross sections as part of the limit setting. Normalisation modifications do also arise from the CP-even counter parts of Eq.~\eqref{eq:opscp} amongst other SMEFT interactions. Hence, limiting ourselves to interference effects provides not only a conservative estimate of the CP-sensitivity reach below, but also targets ``genuine'' CP-violation through designing tell-tale phenomenological discriminators through (ML-generalised) asymmetries. By attributing CP violation to the hard scattering matrix element, we also implicitly assume SM-like decays, hadronisation etc.

We use the NNPDF30NLO (NNPDF23LO) parton distribution functions~\cite{Ball:2012cx} and the default set of parameters that define the \pythia{} setup for parton showering, hadronisation and underlying event activity. For weak-boson fusion and weak-boson scattering processes, it is known that the default parton shower scheme produces too much quark and gluon radiation \cite{ATLAS:2019hoc,Hoche:2021mkv}. For EW $Zjj$ production, which eventually will be analysed in a fiducial region that rejects events with additional jet activity (see Sec.~\ref{sec:limit}), we simulate the SM and interference contributions using the dipole-recoil scheme for the initial-state radiation. For electroweak (EW) $ZZjj$ and EW $W^\pm W^\pm jj$ production, we simulate the SM and interference contributions using the default shower scheme and rescale the sample weights by the ratio $\sigma_\textrm{dipole}^\textrm{SM}/\sigma^\textrm{SM}_\textrm{default}$, where $\sigma^\textrm{SM}$ is the SM cross section in the fiducial region of the analysis and `dipole' or `default' label the shower scheme used to generate the sample. In this latter case, we have checked that the cross section ratio remains approximately flat across the kinematic variables that are of interest in the analysis.

\section{CP-sensitive observables}
\label{sec:cpvar}
\subsection{Simple angular observables}
CP-sensitive observables can be easily constructed using the difference in azimuthal angle between two final particles, i.e.
\begin{equation}
\Delta\phi_{ij} = \phi_i - \phi_j
\end{equation}
where $i$ and $j$ are ordered in rapidity such that $y_i > y_j$. 

The signed azimuthal angle between the jets, $\Delta\phi_{jj}$, is clearly P-odd, and has traditionally been used to search for CP-violation in measurements sensitive to weak boson fusion~\cite{Plehn:2001nj}. We therefore utilise $\Delta\phi_{jj}$ for our analysis of electroweak $Zjj$ production, electroweak $W^\pm W^\pm jj$ production, and electroweak $ZZjj$ production. For the inclusive $W\gamma$, $\gamma\gamma \rightarrow WW$ and electroweak $W^\pm W^\pm jj$ processes, the signed azimuthal angles between charged leptons and photons, $\Delta \phi_{\ell \gamma}$ and $\Delta \phi_{\ell \ell}$, can be constructed as CP-sensitive observables. 

For the electroweak $ZZjj$ process, there are four charged leptons in the final state. CP-violating effects can be probed using the $\Phi_{4\ell{}}$ variable~\cite{Bolognesi:2012mm,Gritsan:2016hjl} defined by
\begin{equation}
\label{eq:cpodd}
    \Phi_{4\ell{}} = \frac{{\bf q}_1 \cdot ( \hat{\bf n}_1 \times \hat{\bf n}_2)}{| {\bf q}_1 \cdot \left( \hat{\bf n}_1 \times \hat{\bf n}_2\right)}| \times {\cos}^{-1}({\bf \hat{\bf n}_1 \cdot \hat{\bf n}_2)}\,,
\end{equation}
which measures the signed angle between the decay planes of the two $Z$ bosons. Here, the normal vectors to the $Z$-boson decay planes are defined as
\begin{equation}
    {\hat{\bf n}_1} = \frac{{\bf q}_{11} \times {\bf q}_{12}}{| {\bf q}_{11} \times {\bf q}_{12}|} \quad {\text{and}} \quad \hat{\bf n}_2 = \frac{{\bf q}_{21} \times {\bf q}_{22}}{|{\bf q}_{21} \times {\bf q}_{22}|},
\end{equation}
where ${\bf q}_{\alpha\beta}$ represents the momentum of the lepton (or antilepton) $\beta$ that originates from the decay $Z_\alpha \to \ell \bar{\ell}$. The ${\bf q}_\alpha = {\bf q}_{\alpha1} + {\bf q}_{\alpha2}$ is the corresponding momentum of the $Z_\alpha$.

\subsection{Observables constructed using neural networks}

CP-sensitive observables can also be constructed using neural networks~\cite{Bhardwaj:2021ujv}. Specifically, for a given interference contribution, the event sample can be divided into positively-weighted and negatively-weighted events. The neural network can then be trained to separate the two classes in a \textit{binary} classification. The SM prediction can be included in the training to simultaneously optimise the separation of the interference contributions from the SM contribution. This is referred to as a \textit{multiclass} model. We mainly focus on the use of multiclass models in this article.

The CP-sensitive observables can be defined on an event-by-event basis using the trained models, i.e. 
\begin{equation}
O_{NN} = P_+ - P_-
\end{equation}
where $P_+$ and $P_-$ are the probabilities that an event is a positively-weighted or negatively-weighted event, respectively. For a binary classification, $P_+ + P_- = 1$. For a multiclass model, $P_+ + P_- + P_{SM} = 1$, where $P_{SM}$ is the probability that an event is a SM event.

We use {\sc{TensorFlow 2.3.0}}~\cite{Abadi:2016kic} to train the neural networks. The input variables are the transverse momentum ($p_\textrm{T}$), pseudorapidity ($\eta$), and azimuthal angle $\phi$ of each visible particle that defines the final state (i.e. leptons, photons, and jets). Lepton flavour and charge are also included, though found to add little sensitivity. In addition, the magnitude ($E_\textrm{T}^\textrm{miss}$) and angle ($\phi_\textrm{miss}$) of the missing transverse momentum are included for  $W\gamma$ production. The choice of hyperparameters is optimised for each network and obtained using the {\sc{Keras}} suite~\cite{keras,omalley2019kerastuner}. The optimisation included the number of layers, the number of nodes, the learning rate and the batch size. 
A data augmentation procedure was used to prevent the networks learning features that arise from statistical fluctuations. In this procedure, each event is used twice in the training, once with the default input variables and once with a CP-operator applied to each of the input variables. The event weight is multiplied by -1 for the CP-flipped events in the interference sample. 

\section{Analysis methodology}
\label{sec:limit}

For each process, we apply the selection cuts used in a recent ATLAS or CMS analysis and, where possible, validate our event generation by comparing the SM fiducial cross section that we obtain from our simulated samples to the theoretical values quoted in the relevant publication. Table~\ref{tab:fiducialregions} presents our fiducial cross sections, which are in good agreement with those reported in the literature after considering differences such as leading-order predictions versus next-to-leading order predictions (in particular in $W\gamma$ production these can be sizeable~\cite{Ohnemus:1992jn,Baur:1993ir}).\footnote{For EW $ZZjj$ production, we only generate the $ZZ\rightarrow 2e2\mu$ final state and our cross section is therefore a factor of two smaller than the theoretical prediction used in Ref.~\cite{ATLAS:2020nlt}.}

\begin{table*}[t!]
    \centering
    \begin{tabular}{|c|c|c|c|}
    \hline 
     Process & Fiducial region & SM cross section & SM yield normalisation \\
     & & & (limit setting only)\\
     \hline 
       EW $Zjj$ ($Z \rightarrow \ell \ell$) & Ref.~\cite{ATLAS:2020nzk} (signal region) & $33.9 \pm 0.1 \textrm{(stat)}$~fb & 3712~\cite{ATLAS:2020nzk}  \\
       inclusive $W\gamma$ ($W\rightarrow \ell \nu$) & Ref.~\cite{CMS:2021cxr} (section 6) & 294~$\pm 1 \textrm{(stat)}$~fb & 102191 \cite{CMS:2021cxr} \\
       EW $ZZjj$ ($ZZ\rightarrow 2e 2\mu)$) &  Ref.~\cite{ATLAS:2020nlt} & $0.09 \pm 0.01 \textrm{(stat)}$~fb & 22.4 \cite{ATLAS:2020nlt} \\
       EW $W^\pm W^\pm jj$ ($W\rightarrow \ell \nu$) & Ref.~\cite{ATLAS:2019cbr} & $2.19 \pm 0.01 \textrm{(stat)}$~fb & 60 \cite{ATLAS:2019cbr}\\
       elastic $\gamma \gamma \rightarrow WW$ ($W\rightarrow \ell \nu$) & Ref.~\cite{ATLAS:2020iwi} (signal region) & $0.67 \pm 0.01 \textrm{(stat)}$~fb & 174 \cite{ATLAS:2020iwi} \\
    \hline   
    \end{tabular}
    \caption{SM fiducial cross sections and event yield normalisations for each of the processes studied in this paper. The fiducial cross sections are calculated within the fiducial region of an ATLAS or CMS experimental analysis and found to be in good agreement with the theoretical predictions reported in those papers. The uncertainties quoted on the fiducial cross sections are statistical and do not contain any systematic uncertainties in the theoretical calculations. The normalisation of the samples is needed for the limit setting and the yields quoted apply to the signal processes. The treatment of backgrounds is discussed in the text.
    \label{tab:fiducialregions}}
\end{table*}

To determine the sensitivity, we construct confidence intervals from each of the CP-odd observables of Sec.~\ref{sec:cpvar} by means of a binned likelihood function 
\begin{equation}
\label{eq:llh}
 {\cal{L}}(\{c_j\}/\Lambda^2) =  \prod_{k} \exp\{-\lambda_k\} \frac{\lambda_k^{n_k}}{n_k!} \,.
\end{equation}
Here, $k$ labels the bins with $n_k$ denoting the expected number of events in bin $k$ assuming the SM-only hypothesis. $\lambda_k$ is the predicted number of events given the SMEFT hypothesis, which is derived from a particular parameter choice in Wilson coefficient space $\{c_j\}/\Lambda^2$. The likelihood of Eq.~\eqref{eq:llh} is converted to a confidence level via a profile-likelihood test statistic~\cite{Feldman:1997qc}. Through Wilks' theorem~\cite{wilks1938}, we can assume this statistic to be distributed according to a $\chi^2$ distribution with one degree of freedom, from which the 95\% confidence level can be obtained. The likelihood function does not include terms that account for systematic uncertainties. This is justified because the constraints are driven by asymmetries in the CP-odd observables, whereas the impact of systematic uncertainties on those distributions would be symmetric. The potential impact of systematic uncertainties is investigated futher in Sec.~\ref{sec:results_syst}.

The $\lambda_k$ and $n_k$ in Eq.~\eqref{eq:llh} are directly derived from the simulated samples, after applying normalisation factors to convert the predicted cross sections into event yields. The normalisation factor is defined for each process such that the SM prediction reproduces the number of SM events that were predicted in the relevant fiducial regions of the ATLAS or CMS analysis (see table~\ref{tab:fiducialregions}). This normalisation factor is applied to both the SM signal prediction and the interference contribution. For the EW $Zjj$ process, the effect of non-EW $Zjj$ backgrounds is included in the calculation of $\lambda_k$ and $n_k$ by generating a QCD $Zjj$ sample 
at ${\cal{O}}(\alpha_s^2 \alpha)$ and normalising that sample to the observed signal-subtracted number of $Zjj$ events observed in Ref.~\cite{ATLAS:2020nzk}. For all other processes, the effects of backgrounds are included in the calculation of $\lambda_k$ and $n_k$ by scaling the SM signal sample to reproduce the number of background events reported in the relevant experimental analysis. This latter approach assumes the kinematic properties of the SM signal and the background are sufficiently similar.

The choice of binning can impact the final confidence level and it is therefore convenient to optimise the binning is such a way that the sensitivity is maximised. Typically about 20 bins are chosen per distribution, with bins subsequently merged to ensure that there is at least one SM event per bin. Post-fit, we check that the theoretical prediction remains positive at the values of the Wilson coefficient that correspond to the 95\% CL, and merge bins if this is not the case.

\section{Results}
\label{sec:results}

\subsection{EW $Zjj$ and inclusive $W\gamma$ production}
\label{sec:results_trip}

\begin{figure}[t!]
\centering
    \includegraphics[width=0.45\textwidth]{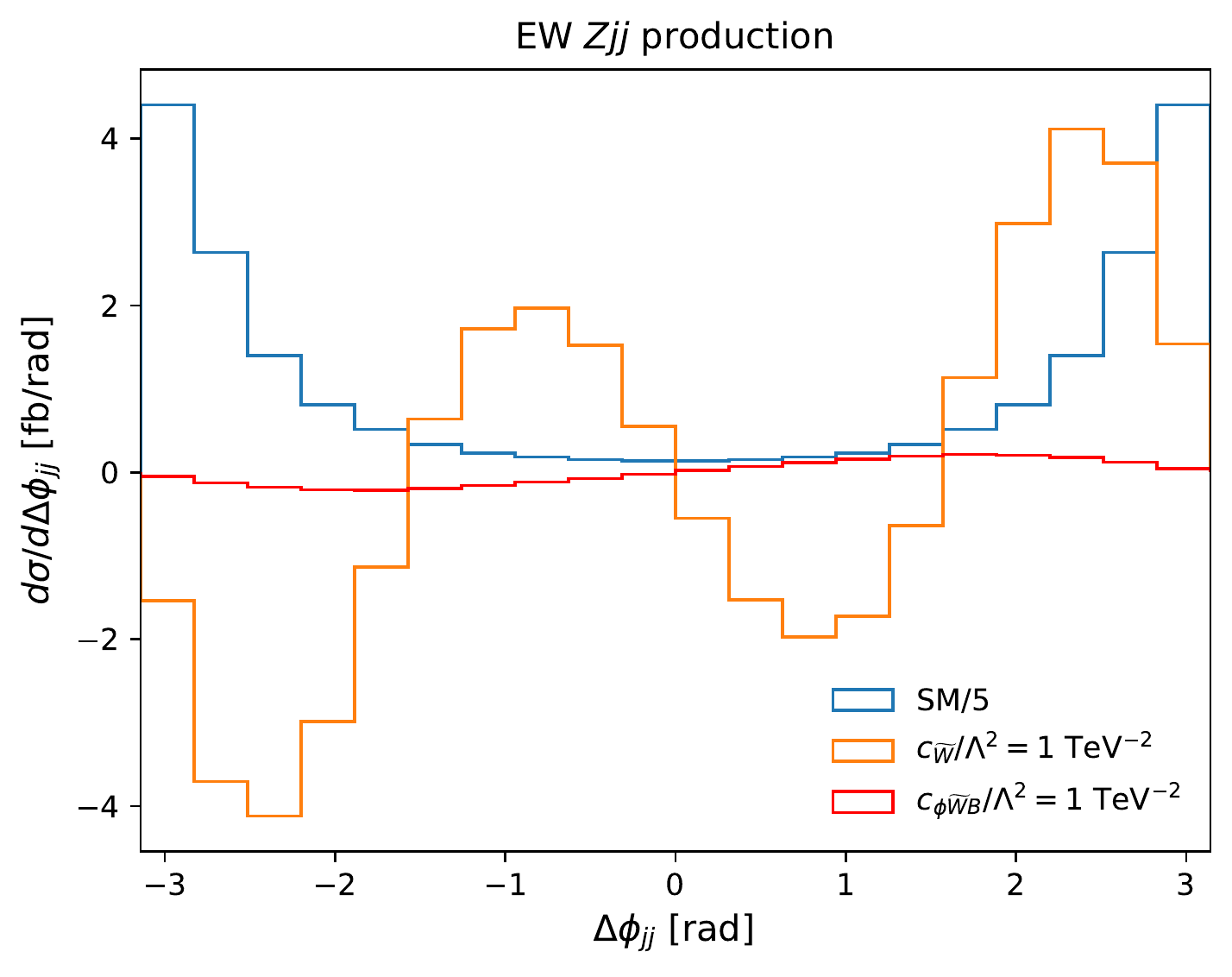}
  \caption{Differential cross sections for the SM and the interference contributions to EW $Zjj$ production as a function of the CP-odd observable $\Delta \phi_{jj}$. The interference contributions are shown for the ${\cal{O}}_{\widetilde{W}}$ and ${\cal{O}}_{\Phi \widetilde{W}B}$ operators.}
  \label{fig:simplecp_zjj}
\end{figure}

\begin{figure}[t!]
\centering
    \includegraphics[width=0.45\textwidth]{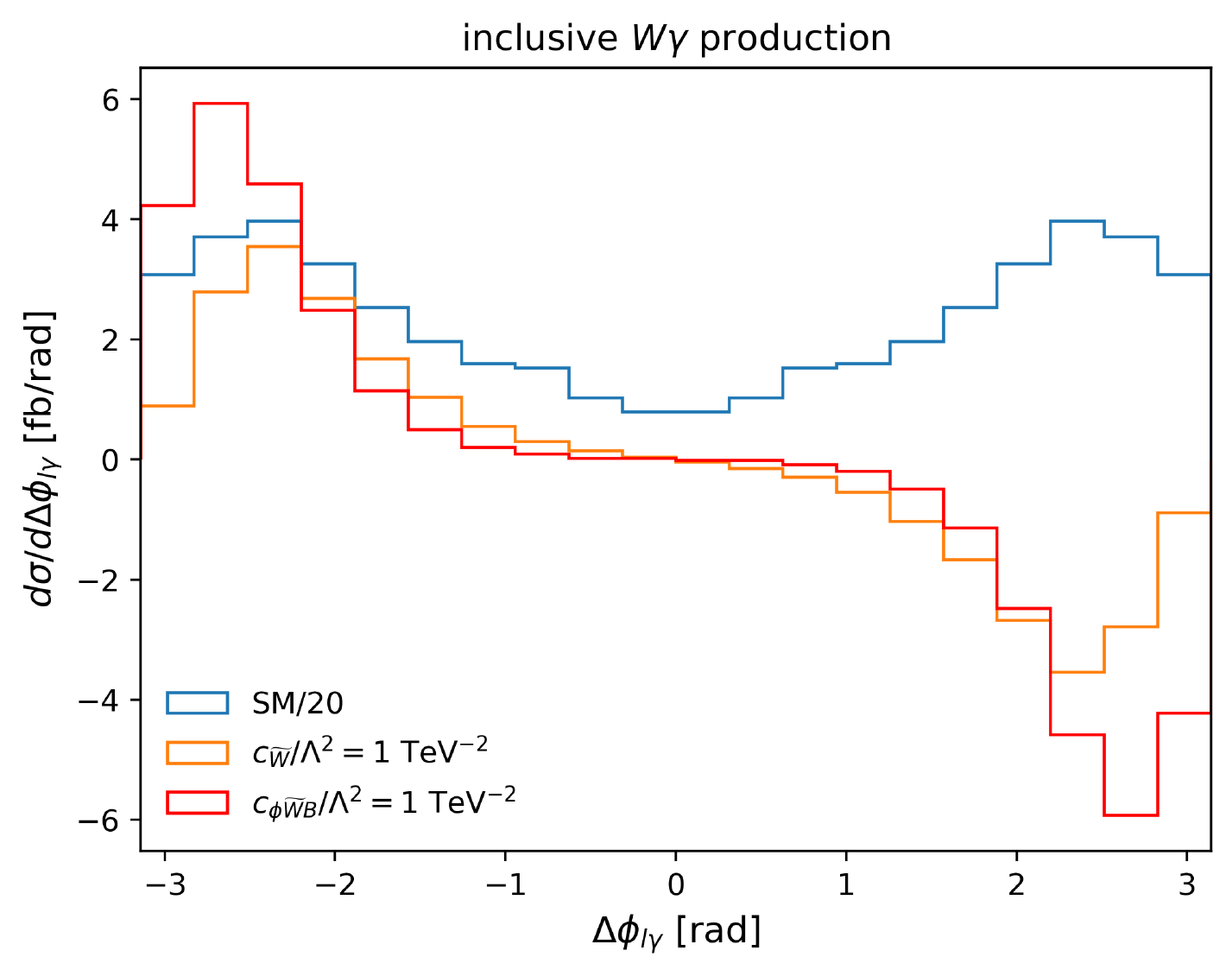}
  \caption{Differential cross sections for the SM and the interference contributions to inclusive $W\gamma$ production as a function of the CP-odd observable $\Delta \phi_{\ell\gamma}$. The interference contributions are shown for the ${\cal{O}}_{\widetilde{W}}$ and ${\cal{O}}_{\Phi \widetilde{W}B}$ operators.}
  \label{fig:simplecp_wy}
\end{figure}

The differential cross section for EW $Zjj$ production as a function of $\Delta \phi_{jj}$  is shown in Fig.~\ref{fig:simplecp_zjj}. Figure~\ref{fig:simplecp_wy} 
shows the differential cross section for inclusive $W\gamma$ production as a function of $\Delta \phi_{\ell \gamma}$. 
In both cases, the SM prediction is shown in addition to the interference contributions induced by the ${\cal{O}}_{\Phi \widetilde{W}B}$ and ${\cal{O}}_{\widetilde{W}}$ operators, with Wilson coefficients set to $c/\Lambda^2=1$~TeV$^{-2}$. As expected, the CP-even SM prediction is symmetric about $\Delta \phi_{jj}$, $\Delta \phi_{\ell \gamma}$ = 0, whereas the CP-odd interference contributions are all asymmetric with an integral of zero. For these simple angular variables, EW $Zjj$ production is most sensitive to the ${\cal{O}}_{\widetilde{W}}$ operator, whereas inclusive $W\gamma$ production is roughly equally sensitive to both operators.

\begin{figure}[t!]
\centering
    \includegraphics[width=0.45\textwidth]{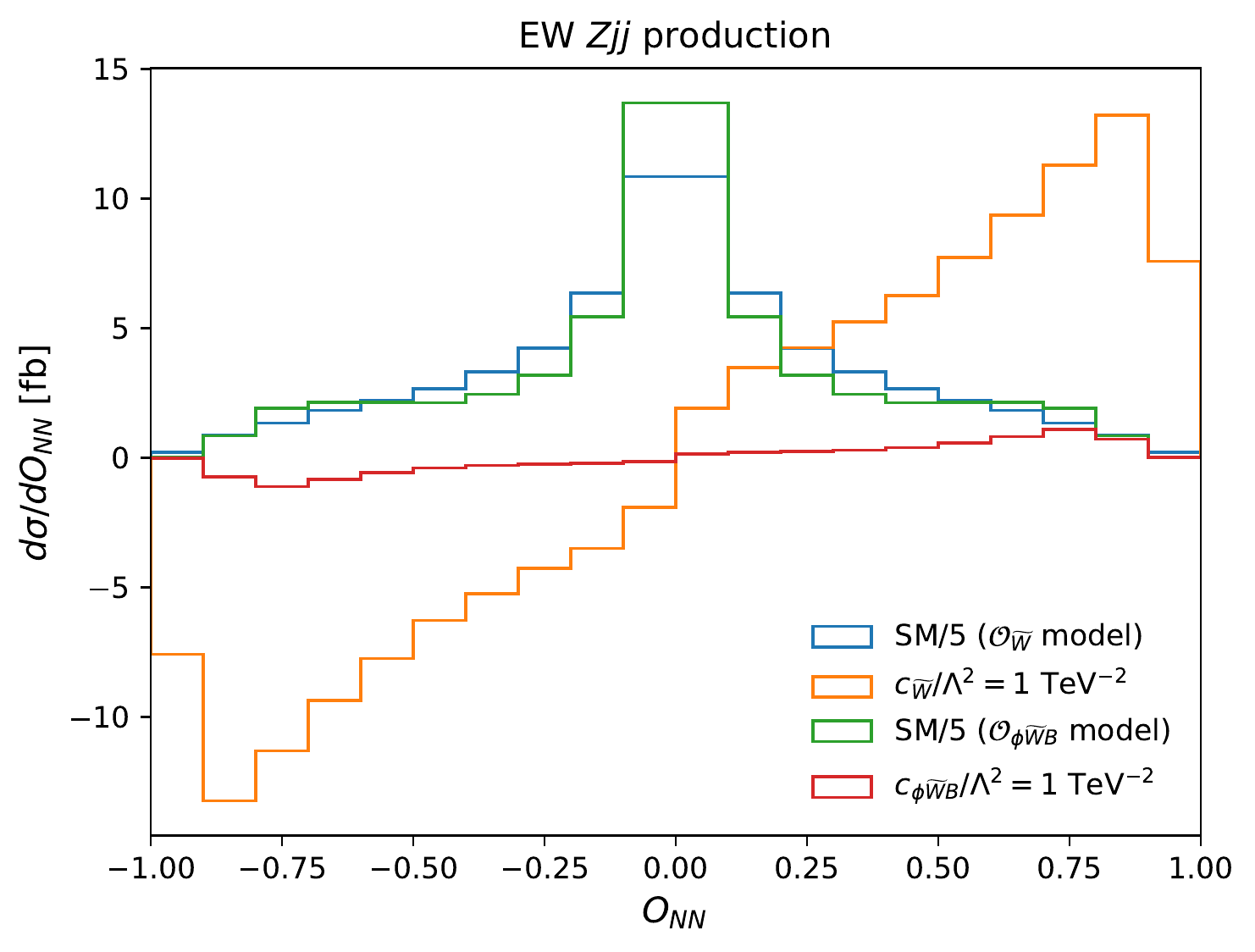}
  \caption{Differential cross sections for the SM and the interference contributions to EW $Zjj$ production as a function of the CP-odd observable $O_{NN}$ (multiclass model). The interference contributions are shown for the ${\cal{O}}_{\widetilde{W}}$ and ${\cal{O}}_{\Phi \widetilde{W}B}$ operators. The neural network was trained separately for each interference contribution.}
  \label{fig:nn_zjj_mult}
\end{figure}

\begin{figure}[t!]
\centering
    \includegraphics[width=0.45\textwidth]{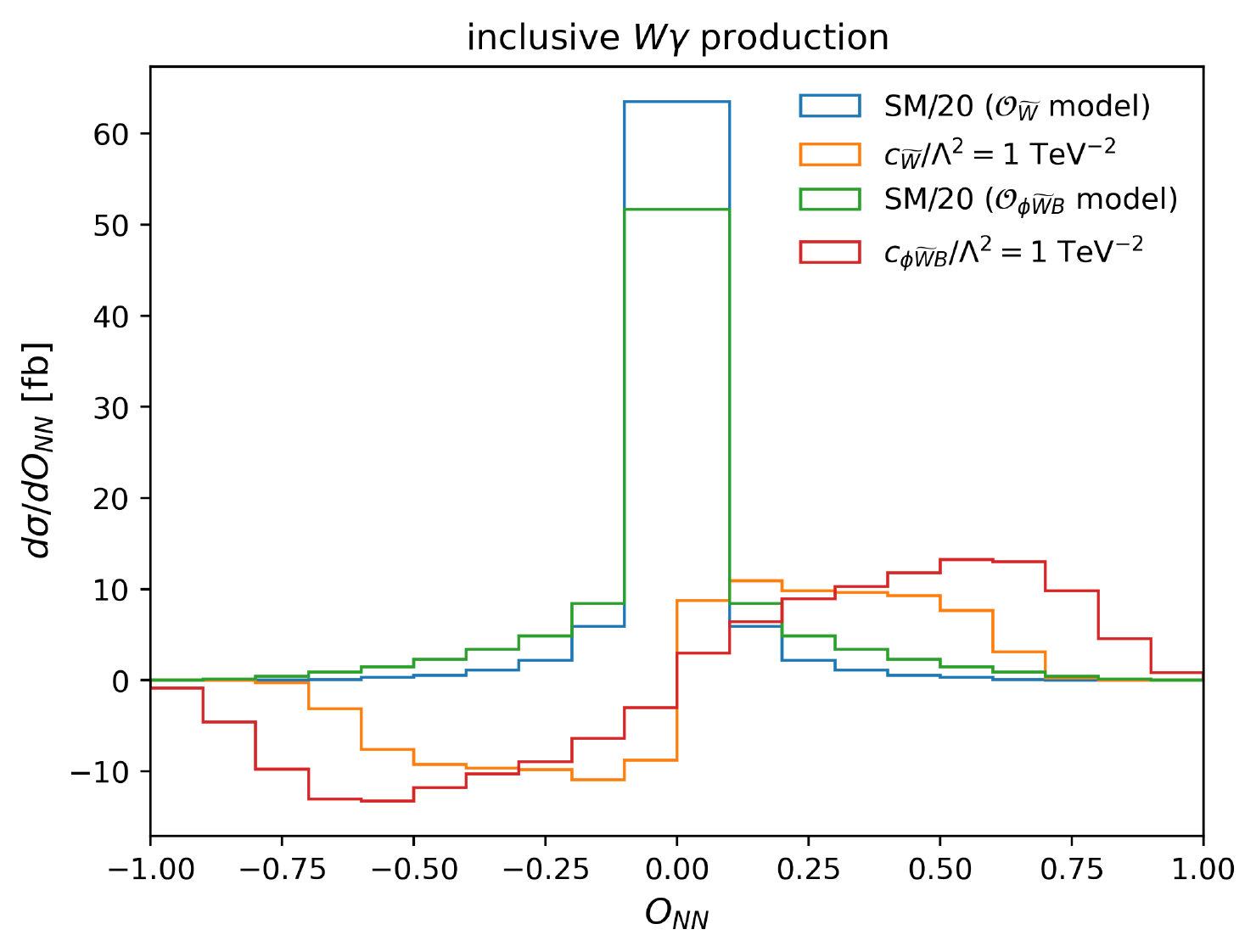}
  \caption{Differential cross sections for the SM and the interference contributions to inclusive $W\gamma$ production as a function of the CP-odd observable $O_{NN}$ (multiclass model). The interference contributions are shown for the ${\cal{O}}_{\widetilde{W}}$ and ${\cal{O}}_{\Phi \widetilde{W}B}$ operators. The neural network was trained separately for each interference contribution.}
  \label{fig:nn_wy_mult}
\end{figure}

The differential cross section for EW $Zjj$ production as a function of the CP-odd observable produced by a multiclass neural network is shown in Fig.~\ref{fig:nn_zjj_mult}. The corresponding distribution for inclusive $W\gamma$ production is shown in Fig.~\ref{fig:nn_wy_mult}. For each operator and process, a neural network is trained to distinguish between the positively-weighted interference contribution, the negatively-weighted interference contribution, and the contribution from the SM. This leads to a different SM distribution depending on the operator being tested. The interference contributions are presented for $c_{\Phi \widetilde{W}B}/\Lambda^2=1$~TeV$^{-2}$ and $c_{\widetilde{W}}/\Lambda^2=1$~TeV$^{-2}$.  

For both processes, the network is effective at separating the positively-weighted and negatively-weighted interference contributions, with the positively-weighted events peaking close to $O_{NN}>0.5$ and the negatively-weighted events at $O_{NN}<-0.5$. The SM contribution is symmetric and peaked at output values very close to zero. In the case of inclusive $W\gamma$ production, it is clear that the network has improved the sensitivity over the use of $\Delta\phi_{\ell \gamma}$ alone, because the interference contributions are much larger relative to the SM prediction.

\begin{table*}[!t]
    \centering
    \begin{tabular}{|c|c|c|c|c|}
    \hline 
     Process & CP-odd observable & $c_{\Phi \widetilde{W}B} / \Lambda^2$~[TeV$^{-2}$] & $c_{ \widetilde{W}} / \Lambda^2$~[TeV$^{-2}$]\\
     \hline 
       & $\Delta\phi_{jj}$ & [-1.05,1.05] & [-0.081,0.081] \\
      EW $Zjj$  & $O_{NN}$ (multi-class)  & [-0.83,0.83] & [-0.047,0.047] \\
& $\Delta\phi_{jj}$ vs $\Delta\phi_{\ell\ell}$  & [-0.99,0.99] & [-0.074,0.074] \\
& $\Delta\phi_{jj}$ vs $p_{\textrm{T},\ell\ell}$  & [-1.04,1.04] & [-0.066,0.066] \\
    \hline   
    & $\Delta\phi_{l\gamma}$ & [-0.165,0.165] & [-0.255,0.255] \\
       inclusive $W\gamma$  &   $O_{NN}$ (multi-class)  & [-0.049,0.049] & [-0.056,0.056] \\
& $\Delta\phi_{l\gamma}$ vs $|\phi_{l} - \phi_\textrm{miss}|$  & [-0.154,0.154] & [-0.219,0.219] \\
& $\Delta\phi_{l\gamma}$ vs $E_\textrm{T}^\textrm{miss}$  & [-0.163,0.163] & [-0.206,0.206] \\
 \hline
 \end{tabular}
    \caption{Expected 95\% confidence interval for the Wilson coefficients affecting triple gauge couplings given an integrated luminosity of 139~fb$^{-1}$ for EW $Zjj$ and inclusive $W\gamma$ production. Results are presented for a one-dimensional fit to the relevant signed-azimuthal angle distribution for each process, as well as fits to the $O_{NN}$ variable constructed from the neural-net outputs of the multi-class models. Each $O_{NN}$ variable is constructed using the interference predicted by the specific operator being tested. 
    \label{tab:vbfz_wy}}
\end{table*}

To quantify the sensitivity of the different observables, we use the profile-likelihood test outlined in Sec.~\ref{sec:limit}. The constraints obtained for each Wilson coefficient are shown in Tab.~\ref{tab:vbfz_wy}. For EW $Zjj$ production, the expected 95\% confidence intervals obtained from fits to the $\Delta\phi_{jj}$ distribution are  similar to those reported by the ATLAS Collaboration~\cite{ATLAS:2020nzk}, with up to 30\% improvement (for $c_{\widetilde{W}} / \Lambda^2$) that is due to (i) using more finely binned distributions\footnote{This is a consequence of the ATLAS measurement being unfolded and requiring more events per bin of the distribution.} and (ii) the small effect of missing systematic uncertainties. For inclusive $W\gamma$ production, the 95\% confidence intervals obtained from fits to the $\Delta\phi_{l\gamma}$ distribution are similar to those reported in Ref.~\cite{Degrande:2021zpv}. The good agreement between our results and those previously reported indicates that our simulation (and subsequent event normalisation) can be used to assess the sensitivity to the different CP-sensitive observables.

The CP-sensitive observables constructed from the output of a multiclass neural network provide much better sensitivity than the simple angular observables alone, with 95\% confidence intervals reduced by a factor of up to two for EW $Zjj$ production and by a factor of 3-5 for inclusive $W\gamma$ production, depending on the specific dimension-six operator. This improvement is similar to that seen in studies of Higgs boson final states in Ref.~\cite{Bhardwaj:2021ujv}. Using the $O_{NN}$ observable, the EW $Zjj$ and inclusive $W\gamma$ processes provide similar sensitivity to CP-violating effects induced by the ${\cal{O}}_{\widetilde{W}}$ operator. The best current experimental constraints on this operator using  linearised EFT contributions are those reported by ATLAS using the $\Delta\phi_{jj}$ distribution in EW $Zjj$ production, i.e. $-0.11<c_{\widetilde{W}}\,(\text{TeV}/\Lambda)^2<0.14$, a factor of two less sensitive. Thus the use of neural networks to construct the CP-sensitive observable and expanding the CP-sensitive measurements to the inclusive $W\gamma$ final state will dramatically improve the sensitivity in the future. For the ${\cal{O}}_{\Phi \widetilde{W}B}$ operator, the constraints obtained from inclusive $W\gamma$ production will be a factor of 17 more precise than can be obtained from measurement of EW $Zjj$ production. This is an important result, because the ATLAS collaboration measured $0.23 < c_{\Phi \widetilde{W}B}\, (\text{TeV}/\Lambda)^2<2.34$ using the $\Delta\phi_{jj}$ distribution in EW $Zjj$ production, with the SM prediction being outside of the 95\% confidence interval.\footnote{This slight asymmetry is not considered to be significant when considering the global $p$-value \cite{ATLAS:2020nzk}. } The measurements of CP-sensitive observables such as $O_{NN}$ for inclusive $W\gamma$ production are therefore critical in searches for CP-violating effects in the Higgs-gauge sector.

\begin{figure}[t!]
\centering
    \includegraphics[width=0.45\textwidth]{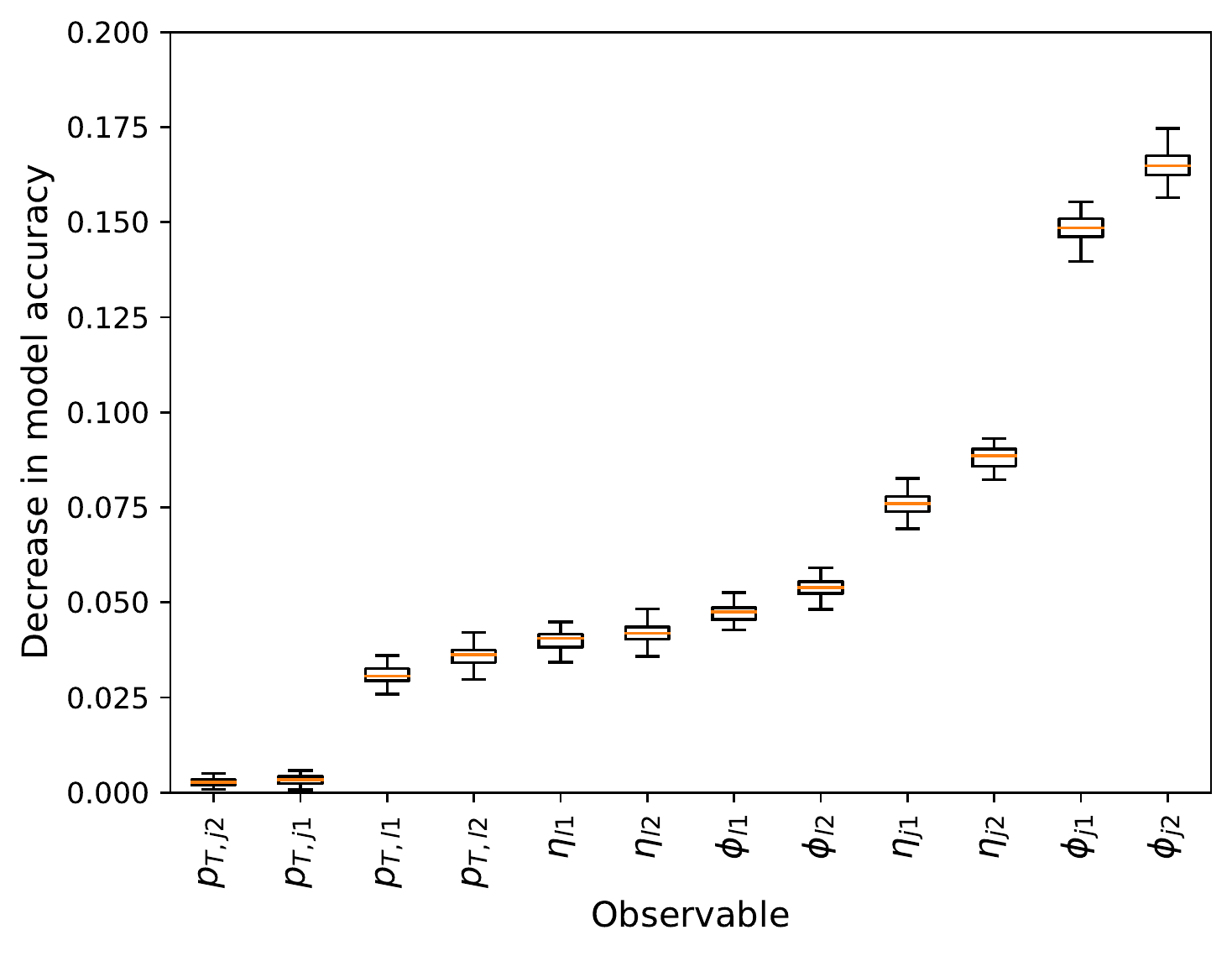}
  \caption{Feature importance for the multiclass network trained on EW $Zjj$ events, with the interference contribution produced by the ${\cal{O}}_{\Phi \widetilde{W}B}$ operator.}
  \label{fig:feat}
\end{figure}

The improved constraints obtained with the CP-sensitive observables constructed using neural networks can be investigated using feature importance techniques. In this approach, the importance of each input variable to the trained network is determined, by evaluating the decrease in accuracy that occurs when the values of the input variable are randomly interchanged in the dataset. An example is shown in Fig.~\ref{fig:feat} for the multiclass network trained on EW $Zjj$ events, with the interference contribution produced by the ${\cal{O}}_{\Phi \widetilde{W}B}$ operator. As expected, the most important variables are the azimuthal angles and rapidities of the two jets, which are needed for the network to learn the $\Delta\phi_{jj}$ observable. However, the azimuthal angles and rapidities of the two leptons are also important, which we trace back to an underlying asymmetry in the $\Delta\phi_{\ell\ell}$ observable, that the network is also exploiting to distinguish between positive and negative interference effects. Finally, the transverse momentum of the two leptons is also important in the multiclass network. However, in our cross check using a binary network, the lepton transverse momentum is not considered to be important. This means that the multiclass network is using the lepton transverse momentum to distinguish between the SM prediction and the interference contributions.


We can use the information from the feature importance tests to construct double-differential distributions that are more sensitive than the simple angular variables alone. The 95\% confidence intervals obtained from (i) a two-dimensional fit to $\Delta\phi_{jj}$ and $\Delta\phi_{\ell\ell}$, as well as (ii) a two-dimensional fit to $\Delta\phi_{jj}$ and $p_{\textrm{T},\ell\ell}$, are also shown in Table.~\ref{tab:vbfz_wy}. For both operators, the constraints are improved using a two-dimensional fit when compared to the use of the simple angular variable alone, but remain inferior to those obtained using a fit to the $O_{NN}$ observable. 

For inclusive $W\gamma$ production, the feature importance tests highlight the importance of the rapidities and azimuthal angles of the lepton and the photon, as expected. However, the azimuthal angle of the missing transverse momentum vector is also found to be important, along with the magnitude of the missing transverse momentum and the transverse momenta of the lepton and photon. The 95\% confidence intervals were determined for a variety of two-dimensional fits to double-differential distributions; the constraints obtained using a two-dimensional fit to $\Delta\phi_{jj}$ and $E_\textrm{T}^\textrm{miss}$ is shown in Table.~\ref{tab:vbfz_wy}, those obtained from a two-dimensional fit to $\Delta\phi_{\ell\gamma}$ and the difference between the azimuthal angles of the lepton and the missing transverse momentum vector. Again, the constraints on the Wilson coefficients are improved with respect to the simple angular variable alone, but do not recover the full sensitivity of the CP-sensitive observable constructed from the output of the multiclass network.

In summary, these two examples demonstrate that CP-sensitive observables constructed using machine-learning techniques can dramatically improve the LHC experiments' sensitivity to CP-violating effects in dimension-six effective field theories.

\subsection{EW $W^\pm W^\pm jj$, EW $ZZjj$ and $\gamma\gamma\rightarrow WW$ production}
\label{sec:results_quart}
In this section, we turn our attention to the processes sensitive to the interaction between four electroweak bosons, namely electroweak $W^\pm W^\pm jj$ production, electroweak $ZZjj$ production and $\gamma\gamma WW$ production. These scattering processes have a significantly reduced SM cross section compared to inclusive $W\gamma$ and EW $Zjj$ production as shown in Tab.~\ref{tab:fiducialregions}, and we can therefore expect these processes to play a less relevant role in constraining the dimension-six operators investigated in this article. Nevertheless, the applicability of machine-learning methods in constructing CP-sensitive observables can still be investigated for each process, which could prove important in the longer term if the LHC experiments start to search for CP-violating effects predicted by dimension-eight effective field theories where multi-gauge boson interactions provide complementary tests to, e.g., $ZZZ$-related interactions (see \cite{Larios:2000ni,Grzadkowski:2016lpv,Belusca-Maito:2017iob,Corbett:2017ecn,Hernandez-Juarez:2021mhi}).

The 95\% confidence intervals obtained for the Wilson coefficients are presented in Tab.~\ref{tab:vbs}, where for each process and operator the constraints obtained using simple angular observables are compared to the constraints obtained using the $O_{NN}$ observable constructed from the output of a multiclass network. These results are obtained by following all of the steps outlined in the previous sections regarding the ML training and limit setting. In general, many of the constraints in Tab.~\ref{tab:vbs} indicate general insensitivity to CP violation as predicted by dimension-six effective field theory and are much less sensitive than those obtained for inclusive $W\gamma$ production and EW $Zjj$ production in Sec.~\ref{sec:results_trip}. However, it is also clear that employing the ML-driven observables leads to significant  improvement in sensitivity when compared to the simple angular observables alone.

\begin{table*}[!t]
    \centering
    \begin{tabular}{|c|c|c|c|c|c|}
    \hline 
     Process & CP-odd observable & $c_{\Phi \widetilde{W}B} / \Lambda^2$~[TeV$^{-2}$] & $c_{ \Phi \widetilde{B}} / \Lambda^2$~[TeV$^{-2}$] & $c_{\Phi \widetilde{W}} / \Lambda^2$~[TeV$^{-2}$] & $c_{\widetilde{W}} / \Lambda^2$~[TeV$^{-2}$]\\
     \hline 
       & $\Delta\phi_{jj}$ & [-3.7,3.7] &  [-43,43] & - & - \\
     EW $ZZjj$ & $\Phi_{4\ell}$ & [-51,51] & [-64,64] & - & -  \\
       &   $O_{NN}$ (multi-class)  & [-3.0,3.0] & [-12,12] & - & - \\

     & $\Delta\phi_{jj}$ & - & - & [-35,34] & [-1.83,1.83] \\
    EW $W^\pm W^\pm jj$ & $\Delta\phi_{\ell\ell}$ & - & - & [-105,105] & [-14,14]  \\
       &   $O_{NN}$ (multi-class)  & - & - & [-17,17] & [-0.76,0.76]  \\
          \hline
    $\gamma\gamma \rightarrow WW$ & $\Delta\phi_{\ell\ell}$ & [-32,32] & [-14,14] & [-48,48] & [-19,19]  \\
       &   $O_{NN}$ (multi-class)  & [-11,11] & [-13,13] & [-43,43] & [-11,11] \\
          \hline      
    \end{tabular}
    \caption{Expected 95\% confidence interval for representative Wilson coefficients given an integrated luminosity of 139~fb$^{-1}$.
     Limits are not competitive in comparison to the trilinear coupling sensitivity of $Zjj$ and $W\gamma$ production. To this end we focus on a motivated subset of operators for $ZZjj$ and $W^\pm W^\pm jj$ production: $Z$-photon mixing highlights $ZZjj$ production as a probe for $B$-like operators as compared to $WWjj$ production. Results are again presented for simple angular observables and for the $O_{NN}$ variable constructed from the outputs of the multi-class networks. Each $O_{NN}$ variable is constructed using the interference predicted by the specific operator being tested. 
    \label{tab:vbs}}
\end{table*}

\begin{figure}[t!]
\centering
    \includegraphics[width=0.45\textwidth]{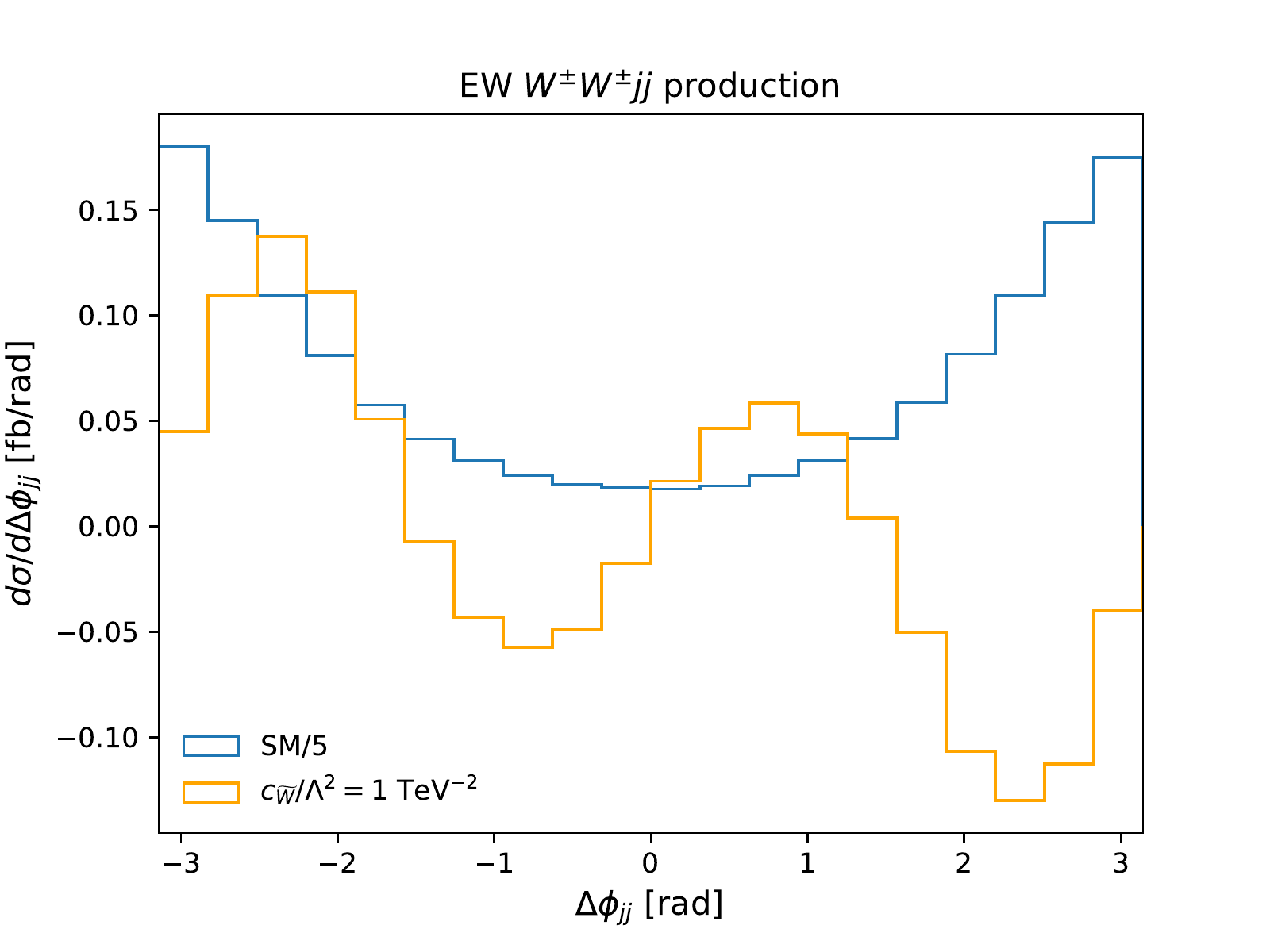}
  \caption{Differential cross sections for the SM and the interference contributions to EW $W^\pm W^\pm jj$ production as a function of the CP-odd observable $\Delta \phi_{jj}$. The interference contribution is shown for the ${\cal{O}}_{\widetilde{W}}$ operator.}
  \label{fig:wwjj_dphijj}
\end{figure}

\begin{figure}[t!]
\centering
    \includegraphics[width=0.45\textwidth]{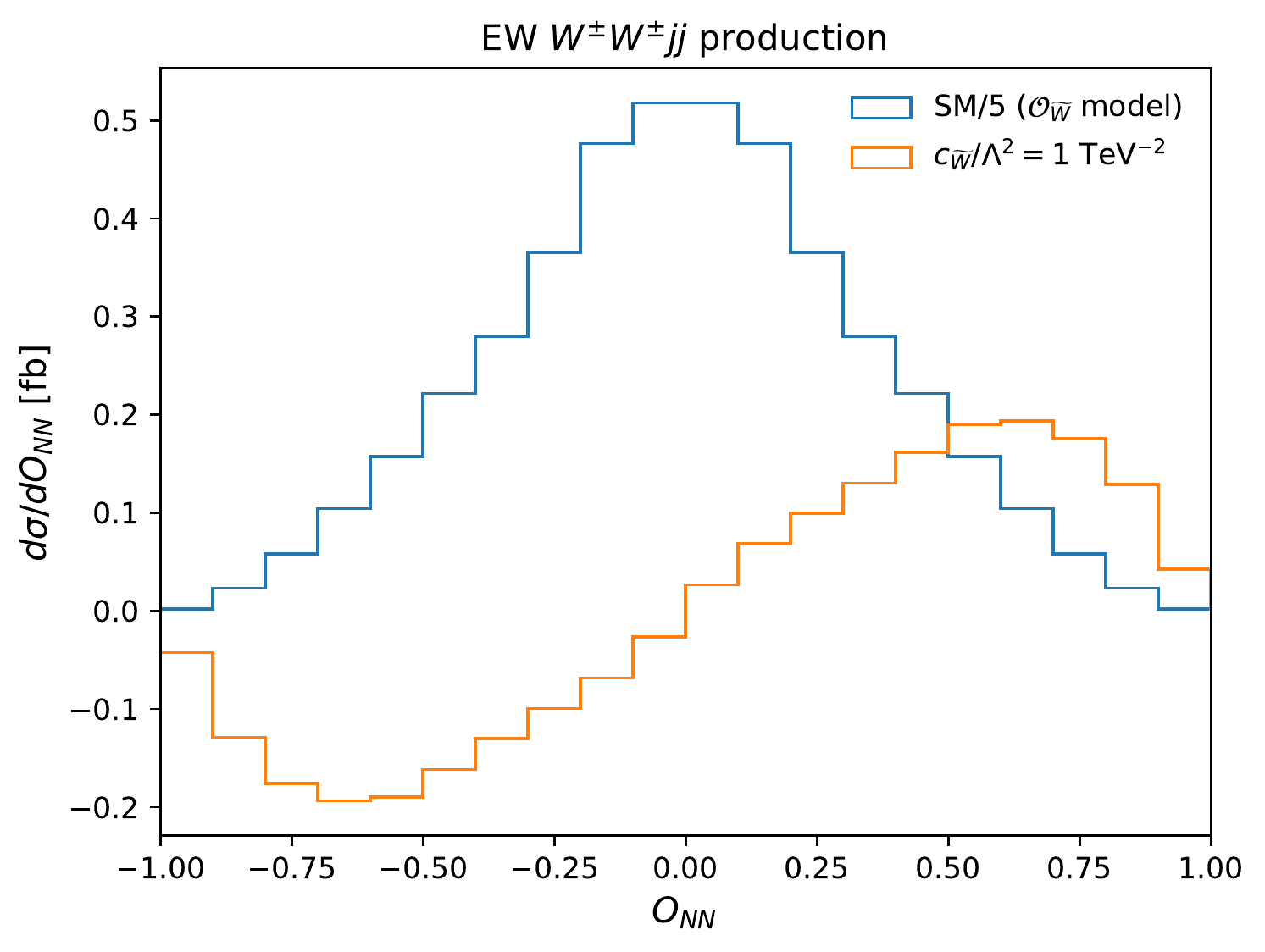}
  \caption{Differential cross sections for the SM and the interference contributions to EW $W^\pm W^\pm jj$ production as a function of the ML-constructed CP-odd observable $O_{NN}$. The interference contribution is shown for the ${\cal{O}}_{\widetilde{W}}$ operator.}
  \label{fig:wwjj_nn}
\end{figure}

In the case of EW $W^\pm W^\pm jj$ production, the $O_{NN}$ observable is sufficiently sensitive to the CP-violating effects predicted  ${\cal{O}}_{\widetilde{W}}$ operator and experimental measurements of $O_{NN}$ for this process would provide useful additional information in a global fit. Unsurprisingly, the sensitivity is mainly driven by the $\Delta \phi_{jj}$ distribution, the differential cross section for which is shown in Fig.~\ref{fig:wwjj_dphijj}. However, the fits to the ML-constructed observable improves the constraints by a factor of about 2.5. Figure~\ref{fig:wwjj_nn} shows the differential cross section as a function of $O_{NN}$, where the asymmetry is clearly enhanced with respect to the SM prediction when compared to the $\Delta\phi_{jj}$ distribution. 

\subsection{Extrapolation to High Luminosity LHC}
\label{sec:results_hllhc}

The constraints obtained on the Wilson coefficients will improve further in the High-Luminosity (HL) phase of the LHC, where the expected integrated luminosity will reach 3~ab$^{-1}$ per experiment. In Table~\ref{tab:hllhc} we present the constraints obtained using the $O_{NN}$ variable constructed from a multiclass model. For these results, the binning was reoptimised, following the procedure outlined in Sec.~\ref{sec:limit} and allowing a larger number of bins to reflect the increased event yields expected at the HL-LHC. In general we find that allowing finer binning does not lead to an  improvement in the 95\% confidence intervals obtained for each Wilson coefficient. The constraints therefore improve by approximately the ratio $(L_\textrm{HL-LHC}/L_\textrm{LHC})^{1/2}$, where $L$ is the integrated luminosity of the dataset (at HL-LHC or LHC as appropriate). This scaling also shows that the sensitivity is dominated by an inclusive selection channeled into
the NN observable, without targeting exclusive phase space regions that become populated at the HL-LHC. Systematic uncertainties of the current LHC runs (see below) will therefore qualitatively carry over to the HL-LHC.

\begin{table*}[!t]
    \centering
    \begin{tabular}{|c|c|c|c|c|}
    \hline 
     Process & $c_{\Phi \widetilde{W}B} / \Lambda^2$~[TeV$^{-2}$] & $c_{ \Phi \widetilde{B}} / \Lambda^2$~[TeV$^{-2}$] & $c_{\Phi \widetilde{W}} / \Lambda^2$~[TeV$^{-2}$] & $c_{\widetilde{W}} / \Lambda^2$~[TeV$^{-2}$]\\
     \hline 
     EW $Zjj$ & [-0.18,0.18] & - & - & [-0.010,0.010] \\
     inclusive $W\gamma$ & [-0.010,0.010] & - & - & [-0.012,0.012] \\
     EW $ZZjj$ & [-0.66,0.66] & [-2.4,2.4] & - & -  \\
     EW $W^\pm W^\pm jj$ & - & - & [-3.9,3.9] & [-0.19,0.19] \\
     $\gamma\gamma\rightarrow WW$ & [-2.5,2.5] & [-2.7,2.7] & [-9.6,9.6] & [-2.5,2.5] \\
          \hline                   
    \end{tabular}
    \caption{Expected 95\% confidence interval for representative Wilson coefficients given an integrated luminosity of 3~ab$^{-1}$ (HL-LHC).
     Results are  presented for the $O_{NN}$ variable. The $O_{NN}$ variable is constructed from a multiclass network that is trained on the interference predicted by the specific operator being tested. Again, we focus on a motivated subset of operators for each production channel. 
    \label{tab:hllhc}}
\end{table*}

\subsection{Impact of systematic uncertainties}
\label{sec:results_syst}

The constraints on the Wilson coefficients presented in the previous sections have been estimated without considering the effect of systematic uncertainties. Although such uncertainties can be large, they will be symmetric as function of a given CP-sensitive observable. In the profile likelihood, therefore, any pull on the associated nuisance parameter to account for asymmetric deviations from the SM prediction would improve the agreement between data and theory in only half of the bins in the distribution, with poorer agreement obtained in the other half of the bins. This should reduce the impact of systematic uncertainties on the 95\% confidence intervals obtained for CP-odd operators.

To demonstrate this effect, we introduce systematic uncertainties in the likelihood function for EW $Zjj$ production and inclusive $W\gamma$ production. For each of these processes, the likelihood function is modified to include a single Gaussian-constrained nuisance parameter, which is intended to account for the systematic uncertainties in the measurements and the theoretical uncertainties in the SM predictions.
The dominant uncertainty for EW $Zjj$ production arises from the theoretical modelling of the non-EW $Zjj$ process, with sizeable contributions also from experimental (jet-energy-scale) uncertainties \cite{vbfzdata}. The total uncertainty is about 15\% uncertainty at low-$|\Delta\phi_{jj}|$ and 25\% at high-$|\Delta\phi_{jj}|$. A weak dependence is observed as a function of the $Z$-boson transverse momentum. For inclusive $W\gamma$ production, the combined experimental and theoretical uncertainty rises from 7\% at low photon transverse momentum to about 25\% at high transverse momentum~\cite{wydata}. However, the uncertainty is relatively flat as a function of angular observables.

The impact of the systematic uncertainties is first tested for EW $Zjj$ production, with the Wilson coefficients constrained using the $|\Delta\phi_{jj}|$ distribution; we find that the 95\% confidence intervals are increased by 1-2\% with respect to those presented in Table~\ref{tab:vbfz_wy}. This is similar to, but slightly smaller than, the effect reported by the ATLAS Collaboration, in which the systematic uncertainties increase the confidence intervals by about 4\% (see the auxiliary material of Ref.~\cite{ATLAS:2020nzk}). It is likely that the smaller impact in our analysis is due to using a simplified model for systematic uncertainties, i.e. that the likelihood is too constraining on the single nuisance parameter. Dropping the Gaussian constraint on the nuisance parameter leads to a slightly larger impact of systematic uncertainties, with the 95\% confidence intervals increased by 2-3\% with respect to Table~\ref{tab:vbfz_wy}. 

The dependence of the systematic uncertainties as a function of $O_{NN}$, for both inclusive EW $Zjj$ production and $W\gamma$ production, is not known and so both uniform and linear dependences are tested in the likelihoods. Again, the impact is small, with a 1\% increase in the confidence intervals if the nuisance parameter has a Gaussian penalty term, and a 2-3\% impact if it does not. From these studies, we conclude that systematic uncertainties will be modest in any future analysis at the LHC.

\section{Conclusions}
\label{sec:conc}
The observed matter-antimatter asymmetry in the Universe requires new physics beyond the Standard Model (SM) and signposts specific areas for phenomenological scrutiny. Specifically, new sources of CP violation are required. Searches for CP violation at the LHC experiments increasingly aim to produce model-independent constraints through the application of effective field theories. Of particular interest is the possibility of CP-violating effects in the Higgs-gauge sector. Although the Higgs sector has received considerable attention along these lines recently, the search for CP violation using processes sensitive to the weak-boson self-interactions is much less explored. These processes however can provide similar sensitivity as the measurements in the Higgs sector, given the gauge structure of the relevant operators in the Standard Model Effective Field Theory (SMEFT). With this in mind, we have investigated a range of purely-electroweak processes in this article, which are sensitive to the underlying nature of the weak-boson self-interactions.

Deviations from the SM are likely to be small, however, showing up predominantly as asymmetries in CP-sensitive observables. Enhancing the new physics reach of the LHC is therefore directly related to formulation of most sensitive BSM discriminators. `Traditional' angular observables, albeit theoretically motivated, might not provide the most sensitive approaches to uncover the anomalous interactions. Many (even termed `optimal') observables have been proposed, typically motivated by considering the angular decompositions of scattering processes. Their choice, however, is not unique. The appropriate choice of observables or observable combinations therefore becomes a question of optimisation that feeds into an enhanced discovery potential at the LHC. This is the natural realm of machine learning (ML), which will perform exactly the task of constructing a tailored CP-sensitive asymmetry optimised to new physics deformations that are parameterised consistently in the SMEFT. 

We have employed the machine-learning techniques proposed in Ref.~\cite{Bhardwaj:2021ujv} to simultaneously construct the CP-sensitive observables and optimise the analysis sensitivity for each of the electroweak processes under investigation. We show that the ML-constructed CP-sensitive observables can lead to large sensitivity enhancements in searches for CP violation at the LHC. The ML algorithm (neural network) achieves this improvement using a multi-class model, which discriminates between the positive- and negative- interference contributions that describe the CP-violating effects, as well as between the interference contributions and the SM prediction. 

We show that future measurements of ML-based CP-sensitive observables for inclusive $W\gamma$ production and EW $Zjj$ production should each improve the sensitivity to CP-violating effects predicted by the  ${\cal{O}}_{\widetilde{W}}$ operator in the effective field theory by a factor of two compared to the current best constraints. We also show that the measurement of ML-based CP-sensitive observables for inclusive $W\gamma$ process can provide a factor of five improvement in sensitivity to the ${\cal{O}}_{\Phi \widetilde{W}B}$ operator over the use of simple angular observables alone. This particular operator in the effective field theory is difficult to constrain; the measurement and technique outlined in this paper is likely to produce the best possible sensitivity.

Although the large improvements in sensitivity are a result of proof-of-principle analyses, they highlight the huge potential that ML-constructed CP-sensitive observables can unleash at the LHC, not only presently with the Run-II and Run-III datasets, but also during the high-luminosity phase in the future.

\bigskip
\noindent{\bf{Acknowledgements}} ---
C.E. is supported by the STFC under grant ST/T000945/1, by the Leverhulme Trust under grant RPG-2021-031, and the IPPP Associateship Scheme. A.D.P is supported by the Royal Society and STFC under grants UF160396, ST/S000925/1 and ST/W000601/1. R.H. and A.D.P are supported by the Leverhulme Trust under grant RPG-2020-004.

\bibliography{main.bbl} 

\providecommand{\href}[2]{#2}\begingroup\raggedright\begin{thebibliography}{10}

\bibitem{ATLAS:2012yve}
{\scshape ATLAS} collaboration, G.~Aad et~al., \emph{{Observation of a new
  particle in the search for the Standard Model Higgs boson with the ATLAS
  detector at the LHC}},
  \href{http://dx.doi.org/10.1016/j.physletb.2012.08.020}{\emph{Phys. Lett. B}
  {\bfseries 716} (2012) 1--29},
  [\href{https://arxiv.org/abs/1207.7214}{{\ttfamily 1207.7214}}].

\bibitem{CMS:2012qbp}
{\scshape CMS} collaboration, S.~Chatrchyan et~al., \emph{{Observation of a New
  Boson at a Mass of 125 GeV with the CMS Experiment at the LHC}},
  \href{http://dx.doi.org/10.1016/j.physletb.2012.08.021}{\emph{Phys. Lett. B}
  {\bfseries 716} (2012) 30--61},
  [\href{https://arxiv.org/abs/1207.7235}{{\ttfamily 1207.7235}}].

\bibitem{Weinberg:1978kz}
S.~Weinberg, \emph{{Phenomenological Lagrangians}},
  \href{http://dx.doi.org/10.1016/0378-4371(79)90223-1}{\emph{Physica A}
  {\bfseries 96} (1979) 327--340}.

\bibitem{Grzadkowski:2010es}
B.~Grzadkowski, M.~Iskrzynski, M.~Misiak and J.~Rosiek, \emph{{Dimension-Six
  Terms in the Standard Model Lagrangian}},
  \href{http://dx.doi.org/10.1007/JHEP10(2010)085}{\emph{JHEP} {\bfseries 10}
  (2010) 085}, [\href{https://arxiv.org/abs/1008.4884}{{\ttfamily 1008.4884}}].

\bibitem{Baak:2014ora}
{\scshape Gfitter Group} collaboration, M.~Baak, J.~C\'uth, J.~Haller,
  A.~Hoecker, R.~Kogler, K.~M\"onig et~al., \emph{{The global electroweak fit
  at NNLO and prospects for the LHC and ILC}},
  \href{http://dx.doi.org/10.1140/epjc/s10052-014-3046-5}{\emph{Eur. Phys. J.
  C} {\bfseries 74} (2014) 3046},
  [\href{https://arxiv.org/abs/1407.3792}{{\ttfamily 1407.3792}}].

\bibitem{Alonso:2013hga}
R.~Alonso, E.~E. Jenkins, A.~V. Manohar and M.~Trott, \emph{{Renormalization
  Group Evolution of the Standard Model Dimension Six Operators III: Gauge
  Coupling Dependence and Phenomenology}},
  \href{http://dx.doi.org/10.1007/JHEP04(2014)159}{\emph{JHEP} {\bfseries 04}
  (2014) 159}, [\href{https://arxiv.org/abs/1312.2014}{{\ttfamily 1312.2014}}].

\bibitem{Grojean:2013kd}
C.~Grojean, E.~E. Jenkins, A.~V. Manohar and M.~Trott, \emph{{Renormalization
  Group Scaling of Higgs Operators and $\Gamma(h \rightarrow \gamma \gamma)$}},
  \href{http://dx.doi.org/10.1007/JHEP04(2013)016}{\emph{JHEP} {\bfseries 04}
  (2013) 016}, [\href{https://arxiv.org/abs/1301.2588}{{\ttfamily 1301.2588}}].

\bibitem{Englert:2014cva}
C.~Englert and M.~Spannowsky, \emph{{Effective Theories and Measurements at
  Colliders}},
  \href{http://dx.doi.org/10.1016/j.physletb.2014.11.035}{\emph{Phys. Lett. B}
  {\bfseries 740} (2015) 8--15},
  [\href{https://arxiv.org/abs/1408.5147}{{\ttfamily 1408.5147}}].

\bibitem{Bakshi:2021ofj}
S.~D. Bakshi, J.~Chakrabortty, C.~Englert, M.~Spannowsky and P.~Stylianou,
  \emph{{Landscaping CP-violating BSM scenarios}},
  \href{http://dx.doi.org/10.1016/j.nuclphysb.2022.115676}{\emph{Nucl. Phys. B}
  {\bfseries 975} (2022) 115676},
  [\href{https://arxiv.org/abs/2103.15861}{{\ttfamily 2103.15861}}].

\bibitem{Naskar:2022rpg}
W.~Naskar, S.~Prakash and S.~U. Rahaman, \emph{{EFT Diagrammatica II: Tracing
  the UV origin of bosonic D6 CPV and D8 SMEFT operators}},
  \href{https://arxiv.org/abs/2205.00910}{{\ttfamily 2205.00910}}.

\bibitem{Degrande:2021zpv}
C.~Degrande and J.~Touch\`eque, \emph{{A Reduced basis for CP violation in
  SMEFT at colliders and its application to Diboson production}},
  \href{https://arxiv.org/abs/2110.02993}{{\ttfamily 2110.02993}}.

\bibitem{ATLAS:2018hxb}
{\scshape ATLAS} collaboration, M.~Aaboud et~al., \emph{{Measurements of Higgs
  boson properties in the diphoton decay channel with 36 fb$^{-1}$ of $pp$
  collision data at $\sqrt{s} = 13$ TeV with the ATLAS detector}},
  \href{http://dx.doi.org/10.1103/PhysRevD.98.052005}{\emph{Phys. Rev. D}
  {\bfseries 98} (2018) 052005},
  [\href{https://arxiv.org/abs/1802.04146}{{\ttfamily 1802.04146}}].

\bibitem{ATLAS:2020wny}
{\scshape ATLAS} collaboration, G.~Aad et~al., \emph{{Measurements of the Higgs
  boson inclusive and differential fiducial cross sections in the 4$\ell$ decay
  channel at $\sqrt{s}$ = 13 TeV}},
  \href{http://dx.doi.org/10.1140/epjc/s10052-020-8223-0}{\emph{Eur. Phys. J.
  C} {\bfseries 80} (2020) 942},
  [\href{https://arxiv.org/abs/2004.03969}{{\ttfamily 2004.03969}}].

\bibitem{ATLAS:2021pkb}
{\scshape ATLAS} collaboration, G.~Aad et~al., \emph{{Constraints on Higgs
  boson properties using $WW^{*}(\rightarrow e\nu\mu\nu) jj$ production in 36.1
  fb$^{-1}$ of $\sqrt{s}$=13 TeV $pp$ collisions with the ATLAS detector}},
  \href{https://arxiv.org/abs/2109.13808}{{\ttfamily 2109.13808}}.

\bibitem{CMS:2021sdq}
{\scshape CMS} collaboration, A.~Tumasyan et~al., \emph{{Analysis of the CP
  structure of the Yukawa coupling between the Higgs boson and $\tau$ leptons
  in proton-proton collisions at $\sqrt{s}$ = 13 TeV}},
  \href{https://arxiv.org/abs/2110.04836}{{\ttfamily 2110.04836}}.

\bibitem{ATLAS:2016ifi}
{\scshape ATLAS} collaboration, G.~Aad et~al., \emph{{Test of CP Invariance in
  vector-boson fusion production of the Higgs boson using the Optimal
  Observable method in the ditau decay channel with the ATLAS detector}},
  \href{http://dx.doi.org/10.1140/epjc/s10052-016-4499-5}{\emph{Eur. Phys. J.
  C} {\bfseries 76} (2016) 658},
  [\href{https://arxiv.org/abs/1602.04516}{{\ttfamily 1602.04516}}].

\bibitem{CMS:2019jdw}
{\scshape CMS} collaboration, A.~M. Sirunyan et~al., \emph{{Constraints on
  anomalous $HVV$ couplings from the production of Higgs bosons decaying to
  $\tau$ lepton pairs}},
  \href{http://dx.doi.org/10.1103/PhysRevD.100.112002}{\emph{Phys. Rev. D}
  {\bfseries 100} (2019) 112002},
  [\href{https://arxiv.org/abs/1903.06973}{{\ttfamily 1903.06973}}].

\bibitem{ATLAS:2020evk}
{\scshape ATLAS} collaboration, G.~Aad et~al., \emph{{Test of CP invariance in
  vector-boson fusion production of the Higgs boson in the $H \rightarrow
  \tau\tau$ channel in proton-proton collisions at $\sqrt{s}=13$ TeV with the
  ATLAS detector}},
  \href{http://dx.doi.org/10.1016/j.physletb.2020.135426}{\emph{Phys. Lett. B}
  {\bfseries 805} (2020) 135426},
  [\href{https://arxiv.org/abs/2002.05315}{{\ttfamily 2002.05315}}].

\bibitem{CMS:2021nnc}
{\scshape CMS} collaboration, A.~M. Sirunyan et~al., \emph{{Constraints on
  anomalous Higgs boson couplings to vector bosons and fermions in its
  production and decay using the four-lepton final state}},
  \href{http://dx.doi.org/10.1103/PhysRevD.104.052004}{\emph{Phys. Rev. D}
  {\bfseries 104} (2021) 052004},
  [\href{https://arxiv.org/abs/2104.12152}{{\ttfamily 2104.12152}}].

\bibitem{Brehmer:2017lrt}
J.~Brehmer, F.~Kling, T.~Plehn and T.~M.~P. Tait, \emph{{Better Higgs-CP Tests
  Through Information Geometry}},
  \href{http://dx.doi.org/10.1103/PhysRevD.97.095017}{\emph{Phys. Rev. D}
  {\bfseries 97} (2018) 095017},
  [\href{https://arxiv.org/abs/1712.02350}{{\ttfamily 1712.02350}}].

\bibitem{Gritsan:2020pib}
A.~V. Gritsan, J.~Roskes, U.~Sarica, M.~Schulze, M.~Xiao and Y.~Zhou,
  \emph{{New features in the JHU generator framework: constraining Higgs boson
  properties from on-shell and off-shell production}},
  \href{http://dx.doi.org/10.1103/PhysRevD.102.056022}{\emph{Phys. Rev. D}
  {\bfseries 102} (2020) 056022},
  [\href{https://arxiv.org/abs/2002.09888}{{\ttfamily 2002.09888}}].

\bibitem{Bortolato:2020zcg}
B.~Bortolato, J.~F. Kamenik, N.~Ko\v{s}nik and A.~Smolkovi\v{c},
  \emph{{Optimized probes of $CP$ -odd effects in the $t \bar t h$ process at
  hadron colliders}},
  \href{http://dx.doi.org/10.1016/j.nuclphysb.2021.115328}{\emph{Nucl. Phys. B}
  {\bfseries 964} (2021) 115328},
  [\href{https://arxiv.org/abs/2006.13110}{{\ttfamily 2006.13110}}].

\bibitem{Barman:2021yfh}
R.~K. Barman, D.~Gon\c{c}alves and F.~Kling, \emph{{Machine Learning the
  Higgs-Top CP Phase}},  \href{https://arxiv.org/abs/2110.07635}{{\ttfamily
  2110.07635}}.

\bibitem{Davis:2021tiv}
J.~Davis, A.~V. Gritsan, L.~S.~M. Guerra, S.~Kyriacou, J.~Roskes and
  M.~Schulze, \emph{{Constraining anomalous Higgs boson couplings to virtual
  photons}},  \href{https://arxiv.org/abs/2109.13363}{{\ttfamily 2109.13363}}.

\bibitem{Bhardwaj:2021ujv}
A.~Bhardwaj, C.~Englert, R.~Hankache and A.~D. Pilkington,
  \emph{{Machine-enhanced CP-asymmetries in the Higgs sector}},
  \href{http://dx.doi.org/10.1016/j.physletb.2022.137246}{\emph{Phys. Lett. B}
  {\bfseries 832} (2022) 137246},
  [\href{https://arxiv.org/abs/2112.05052}{{\ttfamily 2112.05052}}].

\bibitem{ATLAS:2020nzk}
{\scshape ATLAS} collaboration, G.~Aad et~al., \emph{{Differential
  cross-section measurements for the electroweak production of dijets in
  association with a $Z$ boson in proton\textendash{}proton collisions at
  ATLAS}}, \href{http://dx.doi.org/10.1140/epjc/s10052-020-08734-w}{\emph{Eur.
  Phys. J. C} {\bfseries 81} (2021) 163},
  [\href{https://arxiv.org/abs/2006.15458}{{\ttfamily 2006.15458}}].

\bibitem{Plehn:2001nj}
T.~Plehn, D.~L. Rainwater and D.~Zeppenfeld, \emph{{Determining the Structure
  of Higgs Couplings at the LHC}},
  \href{http://dx.doi.org/10.1103/PhysRevLett.88.051801}{\emph{Phys. Rev.
  Lett.} {\bfseries 88} (2002) 051801},
  [\href{https://arxiv.org/abs/hep-ph/0105325}{{\ttfamily hep-ph/0105325}}].

\bibitem{DasBakshi:2020ejz}
S.~Das~Bakshi, J.~Chakrabortty, C.~Englert, M.~Spannowsky and P.~Stylianou,
  \emph{{$CP$ violation at ATLAS in effective field theory}},
  \href{http://dx.doi.org/10.1103/PhysRevD.103.055008}{\emph{Phys. Rev. D}
  {\bfseries 103} (2021) 055008},
  [\href{https://arxiv.org/abs/2009.13394}{{\ttfamily 2009.13394}}].

\bibitem{Biekotter:2021int}
A.~Biek\"otter, P.~Gregg, F.~Krauss and M.~Sch\"onherr, \emph{{Constraining CP
  violating operators in charged and neutral triple gauge couplings}},
  \href{http://dx.doi.org/10.1016/j.physletb.2021.136311}{\emph{Phys. Lett. B}
  {\bfseries 817} (2021) 136311},
  [\href{https://arxiv.org/abs/2102.01115}{{\ttfamily 2102.01115}}].

\bibitem{Alwall:2014hca}
J.~Alwall, R.~Frederix, S.~Frixione, V.~Hirschi, F.~Maltoni, O.~Mattelaer
  et~al., \emph{{The automated computation of tree-level and next-to-leading
  order differential cross sections, and their matching to parton shower
  simulations}}, \href{http://dx.doi.org/10.1007/JHEP07(2014)079}{\emph{JHEP}
  {\bfseries 07} (2014) 079},
  [\href{https://arxiv.org/abs/1405.0301}{{\ttfamily 1405.0301}}].

\bibitem{Sjostrand:2007gs}
T.~Sjostrand, S.~Mrenna and P.~Z. Skands, \emph{{A Brief Introduction to PYTHIA
  8.1}}, \href{http://dx.doi.org/10.1016/j.cpc.2008.01.036}{\emph{Comput. Phys.
  Commun.} {\bfseries 178} (2008) 852--867},
  [\href{https://arxiv.org/abs/0710.3820}{{\ttfamily 0710.3820}}].

\bibitem{Sjostrand:2014zea}
T.~Sj\"ostrand, S.~Ask, J.~R. Christiansen, R.~Corke, N.~Desai, P.~Ilten
  et~al., \emph{{An introduction to PYTHIA 8.2}},
  \href{http://dx.doi.org/10.1016/j.cpc.2015.01.024}{\emph{Comput. Phys.
  Commun.} {\bfseries 191} (2015) 159--177},
  [\href{https://arxiv.org/abs/1410.3012}{{\ttfamily 1410.3012}}].

\bibitem{Brivio:2017btx}
I.~Brivio, Y.~Jiang and M.~Trott, \emph{{The SMEFTsim package, theory and
  tools}}, \href{http://dx.doi.org/10.1007/JHEP12(2017)070}{\emph{JHEP}
  {\bfseries 12} (2017) 070},
  [\href{https://arxiv.org/abs/1709.06492}{{\ttfamily 1709.06492}}].

\bibitem{Brivio:2020onw}
I.~Brivio, \emph{{SMEFTsim 3.0 \textemdash{} a practical guide}},
  \href{http://dx.doi.org/10.1007/JHEP04(2021)073}{\emph{JHEP} {\bfseries 04}
  (2021) 073}, [\href{https://arxiv.org/abs/2012.11343}{{\ttfamily
  2012.11343}}].

\bibitem{Degrande:2011ua}
C.~Degrande, C.~Duhr, B.~Fuks, D.~Grellscheid, O.~Mattelaer and T.~Reiter,
  \emph{{UFO - The Universal FeynRules Output}},
  \href{http://dx.doi.org/10.1016/j.cpc.2012.01.022}{\emph{Comput. Phys.
  Commun.} {\bfseries 183} (2012) 1201--1214},
  [\href{https://arxiv.org/abs/1108.2040}{{\ttfamily 1108.2040}}].

\bibitem{Ball:2012cx}
R.~D. Ball et~al., \emph{{Parton distributions with LHC data}},
  \href{http://dx.doi.org/10.1016/j.nuclphysb.2012.10.003}{\emph{Nucl. Phys. B}
  {\bfseries 867} (2013) 244--289},
  [\href{https://arxiv.org/abs/1207.1303}{{\ttfamily 1207.1303}}].

\bibitem{ATLAS:2019hoc}
{\scshape ATLAS} collaboration, G.~Aad et~al., \emph{{Modelling of the vector
  boson scattering process $pp\rightarrow W^\pm W^\pm jj$ in Monte Carlo
  generators in ATLAS}}, {\emph{ATL-PHYS-PUB-2019-004} (2019) }.

\bibitem{Hoche:2021mkv}
S.~H\"oche, S.~Mrenna, S.~Payne, C.~T. Preuss and P.~Skands, \emph{{A Study of
  QCD Radiation in VBF Higgs Production with Vincia and Pythia}},
  \href{http://dx.doi.org/10.21468/SciPostPhys.12.1.010}{\emph{SciPost Phys.}
  {\bfseries 12} (2022) 010},
  [\href{https://arxiv.org/abs/2106.10987}{{\ttfamily 2106.10987}}].

\bibitem{Bolognesi:2012mm}
S.~Bolognesi, Y.~Gao, A.~V. Gritsan, K.~Melnikov, M.~Schulze, N.~V. Tran
  et~al., \emph{{On the spin and parity of a single-produced resonance at the
  LHC}}, \href{http://dx.doi.org/10.1103/PhysRevD.86.095031}{\emph{Phys. Rev.
  D} {\bfseries 86} (2012) 095031},
  [\href{https://arxiv.org/abs/1208.4018}{{\ttfamily 1208.4018}}].

\bibitem{Gritsan:2016hjl}
A.~V. Gritsan, R.~R\"ontsch, M.~Schulze and M.~Xiao, \emph{{Constraining
  anomalous Higgs boson couplings to the heavy flavor fermions using matrix
  element techniques}},
  \href{http://dx.doi.org/10.1103/PhysRevD.94.055023}{\emph{Phys. Rev. D}
  {\bfseries 94} (2016) 055023},
  [\href{https://arxiv.org/abs/1606.03107}{{\ttfamily 1606.03107}}].

\bibitem{Abadi:2016kic}
M.~Abadi et~al., \emph{{TensorFlow: Large-Scale Machine Learning on
  Heterogeneous Distributed Systems}},
  \href{https://arxiv.org/abs/1603.04467}{{\ttfamily 1603.04467}}.

\bibitem{keras}
F.~Chollet, ``Keras.'' \url{https://github.com/fchollet/keras}, 2015.

\bibitem{omalley2019kerastuner}
T.~O'Malley, E.~Bursztein, J.~Long, F.~Chollet, H.~Jin, L.~Invernizzi et~al.,
  ``Kerastuner.'' \url{https://github.com/keras-team/keras-tuner}, 2019.

\bibitem{Ohnemus:1992jn}
J.~Ohnemus, \emph{{Order-$\alpha_s$ calculations of hadronic $W^\pm \gamma$ and
  $Z \gamma$ production}},
  \href{http://dx.doi.org/10.1103/PhysRevD.47.940}{\emph{Phys. Rev. D}
  {\bfseries 47} (1993) 940--955}.

\bibitem{Baur:1993ir}
U.~Baur, T.~Han and J.~Ohnemus, \emph{{QCD corrections to hadronic $W \gamma$
  production with nonstandard $W W \gamma$ couplings}},
  \href{http://dx.doi.org/10.1103/PhysRevD.48.5140}{\emph{Phys. Rev. D}
  {\bfseries 48} (1993) 5140--5161},
  [\href{https://arxiv.org/abs/hep-ph/9305314}{{\ttfamily hep-ph/9305314}}].

\bibitem{ATLAS:2020nlt}
{\scshape ATLAS} collaboration, G.~Aad et~al., \emph{{Observation of
  electroweak production of two jets and a $Z$-boson pair}},
  \href{https://arxiv.org/abs/2004.10612}{{\ttfamily 2004.10612}}.

\bibitem{CMS:2021cxr}
{\scshape CMS} collaboration, A.~Tumasyan et~al., \emph{{Measurement of
  W$^\pm\gamma$ differential cross sections in proton-proton collisions at
  $\sqrt{s}$ = 13 TeV and effective field theory constraints}},
  \href{http://dx.doi.org/10.1103/PhysRevD.105.052003}{\emph{Phys. Rev. D}
  {\bfseries 105} (2022) 052003},
  [\href{https://arxiv.org/abs/2111.13948}{{\ttfamily 2111.13948}}].

\bibitem{ATLAS:2019cbr}
{\scshape ATLAS} collaboration, M.~Aaboud et~al., \emph{{Observation of
  electroweak production of a same-sign $W$ boson pair in association with two
  jets in $pp$ collisions at $\sqrt{s}=13$ TeV with the ATLAS detector}},
  \href{http://dx.doi.org/10.1103/PhysRevLett.123.161801}{\emph{Phys. Rev.
  Lett.} {\bfseries 123} (2019) 161801},
  [\href{https://arxiv.org/abs/1906.03203}{{\ttfamily 1906.03203}}].

\bibitem{ATLAS:2020iwi}
{\scshape ATLAS} collaboration, G.~Aad et~al., \emph{{Observation of
  photon-induced $W^+W^-$ production in $pp$ collisions at $\sqrt{s}=13$ TeV
  using the ATLAS detector}},
  \href{http://dx.doi.org/10.1016/j.physletb.2021.136190}{\emph{Phys. Lett. B}
  {\bfseries 816} (2021) 136190},
  [\href{https://arxiv.org/abs/2010.04019}{{\ttfamily 2010.04019}}].

\bibitem{Feldman:1997qc}
G.~J. Feldman and R.~D. Cousins, \emph{{A Unified approach to the classical
  statistical analysis of small signals}},
  \href{http://dx.doi.org/10.1103/PhysRevD.57.3873}{\emph{Phys. Rev. D}
  {\bfseries 57} (1998) 3873--3889},
  [\href{https://arxiv.org/abs/physics/9711021}{{\ttfamily physics/9711021}}].

\bibitem{wilks1938}
S.~S. Wilks, \emph{{The Large-Sample Distribution of the Likelihood Ratio for
  Testing Composite Hypotheses}},
  \href{http://dx.doi.org/10.1214/aoms/1177732360}{\emph{Annals Math. Statist.}
  {\bfseries 9} (1938) 60--62}.

\bibitem{Larios:2000ni}
F.~Larios, M.~A. Perez, G.~Tavares-Velasco and J.~J. Toscano, \emph{{Trilinear
  neutral gauge boson couplings in effective theories}},
  \href{http://dx.doi.org/10.1103/PhysRevD.63.113014}{\emph{Phys. Rev. D}
  {\bfseries 63} (2001) 113014},
  [\href{https://arxiv.org/abs/hep-ph/0012180}{{\ttfamily hep-ph/0012180}}].

\bibitem{Grzadkowski:2016lpv}
B.~Grzadkowski, O.~M. Ogreid and P.~Osland, \emph{{CP-Violation in the $ZZZ$
  and $ZWW$ vertices at $e^+e^-$ colliders in Two-Higgs-Doublet Models}},
  \href{http://dx.doi.org/10.1007/JHEP05(2016)025}{\emph{JHEP} {\bfseries 05}
  (2016) 025}, [\href{https://arxiv.org/abs/1603.01388}{{\ttfamily
  1603.01388}}].

\bibitem{Belusca-Maito:2017iob}
H.~B\'elusca-Ma\"\i{}to, A.~Falkowski, D.~Fontes, J.~C. Rom\~ao and J.~a.~P.
  Silva, \emph{{CP violation in 2HDM and EFT: the $ZZZ$ vertex}},
  \href{http://dx.doi.org/10.1007/JHEP04(2018)002}{\emph{JHEP} {\bfseries 04}
  (2018) 002}, [\href{https://arxiv.org/abs/1710.05563}{{\ttfamily
  1710.05563}}].

\bibitem{Corbett:2017ecn}
T.~Corbett, M.~J. Dolan, C.~Englert and K.~Nordstr\"om, \emph{{Anomalous
  neutral gauge boson interactions and simplified models}},
  \href{http://dx.doi.org/10.1103/PhysRevD.97.115040}{\emph{Phys. Rev. D}
  {\bfseries 97} (2018) 115040},
  [\href{https://arxiv.org/abs/1710.07530}{{\ttfamily 1710.07530}}].

\bibitem{Hernandez-Juarez:2021mhi}
A.~I. Hern\'andez-Ju\'arez, A.~Moyotl and G.~Tavares-Velasco,
  \emph{{Contributions to $ZZV^*$ ($V=\gamma ,Z,Z'$) couplings from $CP$
  violating flavor changing couplings}},
  \href{http://dx.doi.org/10.1140/epjc/s10052-021-09093-w}{\emph{Eur. Phys. J.
  C} {\bfseries 81} (2021) 304},
  [\href{https://arxiv.org/abs/2102.02197}{{\ttfamily 2102.02197}}].

\bibitem{vbfzdata}
{\scshape ATLAS} collaboration.
\newblock https://doi.org/10.17182/hepdata.94218.

\bibitem{wydata}
{\scshape CMS} collaboration.
\newblock https://doi.org/10.17182/hepdata.115354.

\end{thebibliography}\endgroup

\end{document}